# Effect of geometrical, operational and material parameters in the lubrication regime of hard-on-hard hip implants

K. Nitish Prasad, P. Ramkumar*

Advanced Tribology Research Lab (ATRL), Machine Design Section, Department of Mechanical Engineering,

Indian Institute of Technology Madras (IITM), Chennai, India.

*Corresponding author: ramkumar@iitm.ac.in

**Abstract**

Several factors, such as the material, geometry, roughness, relative angular movement and loading conditions, influence the hip implant's lubrication regime, thereby affecting the friction and wear performance including longevity. In this study, a ball-on-plane model is considered to predict the lubrication regime across the entire gait cycle for the metal-on-metal and ceramic-on-ceramic hip implants. According to ISO 14242-1, a normal walking gait cycle is considered for the loads and rotations in the hip implant for this study. A comprehensive analysis is done to estimate the lubrication regime by considering different material combinations, body weights, femoral head sizes, clearances, and roughness across the entire gait cycle. The correlation coefficients show that the geometrical parameters are dominant in affecting the lubrication film thickness compared to the operational and material parameters. The femoral head size and body weight are found to be the most dominant and least dominant parameter respectively. Among the hard-on-hard tribo-pair, ceramics operate predominantly in the full-film lubrication regime compared to metals due to its fine surface finish. The present research suggests that in order to maximise the longevity of a hip implant, an orthopaedic surgeon should choose one with a larger femoral head diameter, less clearance, and an ultra-fine surface finish, regardless of any tribo-pair.

## 1.Introduction

The natural hip joint represents a ball-on-socket configuration providing three rotational degrees of freedom, resulting in human gait motion. When the natural hip joint degenerates, it is replaced by Total Hip Replacement (THR) and Resurfacing Hip Replacement (HRR) to aid gait movement based on the patient's requirements. Different material combinations have been employed in THR and HRR tribo-pairs, such as the hard-on-soft and hard-on-hard configurations [1]. It is reported that the hard-on-hard material combination, especially Metal-on-Metal (MoM) and Ceramic-on-Ceramic (CoC), offer high wear resistance which is suitable for younger patients improving the implant's longevity [2,3]. Even though the hard-on-hard hip implants are tested in vitro through critical ISO standards, failures occur in vivo due to aseptic loosening and fracture of the implant [4]. The significant contribution of these typical failures comes from friction and wear occurring in articulating the tribo-pair (acetabular cup and femoral head) [5,6]. Lubrication in the contact surface needs to be investigated post-surgery, which affects the tribo-performance of the implant. The thickness of the lubricant film denotes the type of regime in which the hip implant operates; a very-thin film lesser than the surface roughness where direct contact occurs between the contact surfaces (boundary lubrication); a thin film that could not separate all the asperity contact points between the contact surfaces (mixed lubrication); a sufficiently thick film which can completely separate the contact surfaces (full-film lubrication). As friction and wear are system properties, their values will be affected due to several parameters such as material combination, loads, velocity, environment, etc. Therefore, it is essential to investigate the lubricant performance of the hip implant under various geometric, operational and material parameters.

Researchers have used in-vitro and in-vivo methods to estimate the film formation and lubrication performance in hip implants. The friction values in hip implants are generally measured through pendulum simulators with a static or dynamic load [7,8]. The mode of lubrication is determined based on the friction values obtained from the simulator. Some studies [9,10] have estimated the lubricant film gap by optical interferometry images. However, gaps from images do not signify the actual lubricant film thickness under all implant working conditions. Wear performance is estimated in-vitro through hip simulators or simple pin-on-plate or pin-on-disc tests [11–13]. Still, the wear results from implants obtained in-vivo seem different from the ones obtained using hip simulators. Because in-vivo, other complications such as surgical malpositioning, patient anatomy, lubricant rheological properties, and critical loading conditions would affect the friction and wear performance of the implant. Experiments with complex equipment that is costly, time-consuming, and labour-intensive are needed to test all of these conditions.

Numerous lubricant simulation models [14–16] have been found to study the lubricant performance of the hip implant by measuring the lubricant film thickness. Hamrock and Dowson developed an analytical model to study the lubrication regime with respect to point contacts based on the system parameters. Studies [16] have used a general Reynold's equation to model the minimum film thickness in hard-on-soft hip implants. The high deformation in UHMWPE polymers makes the theory of a semi-infinite space to be invalid. However, it should be noted that the influence of the cup inclination angle is negligible up to 45°, and the flexion-extension is only considered as the dominant motion affecting the minimum film thickness.

Wang et al. [17] studied the effect of lubrication in metal-on-metal hip implants using both ball-on-socket and ball-on-plane models. The authors compared the results of both models and established that consideration of the ball-on-plane model is valid unless the contact is at the edge of the cup surface (high inclination angle > 60°). Mattei et al. [18] did EHL modelling of hip implants under normal walking quasi-static conditions for both hard-on-soft and hard-on-hard configuration. The authors concluded that the ball-on-plane assumption holds well for the hard-on-hard hip implants compared to the hard-on-soft hip implants. Most of the studies performed steady-state EHL analysis to determine the lubricant film thickness over a domain on the contact surface with a fixed head size and clearance. However, the present study aims to evaluate only the minimum film thickness (which occurs at a single point in a domain) among the whole region of contact using different operational, geometric and material parameters. Even though the considered parameters for analysis change with respect to time, consideration of initial parameters will help to determine the lubrication performance during the initial running-in period of the application where the effect of wear is high. Further, prediction of the lubrication regime during the initial condition will help in understanding the shift of the lubrication regime for the prolonged life of the implant.

In the present study, the lubrication performance of the hard-on-hard hip implant combinations such as MoM and CoC will be investigated under a dynamic normal walking cycle using the Hamrock-Dowson equation considering the parameters stated below:

a) Geometrical parameters (controllable): femoral head diameter, clearance and surface roughness
b) Operational parameters (uncontrollable): body weight of the patient, gait pattern force and articulating velocity
c) Material parameters (controllable): Young's modulus and Poisson's ratio

The parameters are further grouped into controllable and uncontrollable, depending on whether the orthopaedic surgeon has control over the decision about the hip implant prior to surgery or if the patient's behaviour is beyond the doctor's influence. The outcome of the study will determine the critical parameters affecting the lubricant performance and operating lubrication regimes of the hard-on-hard hip prostheses under the grouped parameters. Further, the individual effect of these parameters is studied and rated to affect the performance of the artificial joint. The results of the study are expected to help the orthopaedist in choosing the appropriate hip implant based on the controllable parameters. Overall, the present research helps in selecting the best hip implant for a prolonged life.

## 2. Materials and Methodology

### 2.1 Materials and geometrical parameters

The most commonly used hard-on-hard combinations are MoM and CoC, which are analysed in the present study. The material properties of the hard-on-hard combinations considered in this study are listed in Table 1.

**Table 1**

Input material properties

| Tribo-pair | Material | Young's Modulus, E (GPa) | Poisson's ratio, $v$ |
|---|---|---|---|
| MoM | CoCrMo | 210 | 0.30 |
| CoC | ZTA (Biolox Delta) | 350 | 0.26 |

Despite the concerns regarding the metal-ion release, CoCrMo alloy is still widely used as the material in tribo-pair due to biocompatibility, strength, durability and wear resistance compared to other metallic alloys [2,3,19,20]. So, CoCrMo is considered as the material for MoM tribo-pair in this study. A composite material made of Zirconia and Alumina called Zirconia Toughened Alumina (ZTA- commercially named "Biolox Delta") is considered as the ceramic material for the CoC tribo-pair. This ceramic material provides high fracture strength, bio-inert wear debris and elevated resistance to crack propagation compared to other existing bio-ceramic materials [3,21–24]. The geometrical and operational parameters which are considered for the analysis in the present study are tabulated in Table 2.

**Table 2**

Input geometrical and operational parameters

| Tribo-pair | MoM and CoC |
|---|---|
| Femoral head diameter $D$ (*mm*) | 28, 32, 36 (THR cases) and 44, 58 (HRR cases) |
| Radial clearance $C$ (*μm*) | 25, 50, 75, 100 |
| Body weight $BW$ (*kg*) | 60, 80, 100, 120 |
| Viscosity of the body fluid (*Pa.s*) | 0.0025 |
| Surface Roughness $R_a$ (*μm*) | 0.03 for MoM and 0.003 for CoC |

Three different femoral head sizes of diameter (28,32, and 36 mm) are considered with the application related to THR [25] , and two head sizes of diameter (44 and 58 mm) are considered with the application associated with HRR [26]. The particular head sizes and clearances are selected to cover a wide range of patients and implant manufacturers in terms of demography, stability, patient anatomy and surgeon preference. Eq. (1) provides a mathematical representation of the radial clearance C, which is computed as the difference between the nominal radius of the head and the cup, as illustrated in Fig. 1a.

$$C = R_{cup} - R_{head} \tag{1}$$

## 2.2 Ball-on-plane Geometric configuration

On the contrary to the ball-on-socket model in hard-on-hard implants, a ball-on-plane model shown in Fig. 1a is taken into consideration in predicting the lubricating behaviour, as reported [18,27,28]. The only limitation of the ball-on-plane model is the accuracy of the prediction unless the head is in contact with the cup at a higher inclination angle [17]. However, in this study, contact is assumed to occur only inside the bearing surface of the cup with a 45° inclination angle. Also, as the hard-on-hard combination is subjected to a very less contact area in comparison to the cup radius, thus the assumption of a semi-infinite space is valid [27] , as shown in Fig. 1b. The equivalent radius $R_{eq}$ can be predicted for the ball-on-plane model as shown in Eq.(2).

$$R_{eq} = \frac{R_{cup} R_{head}}{C} \tag{2}$$

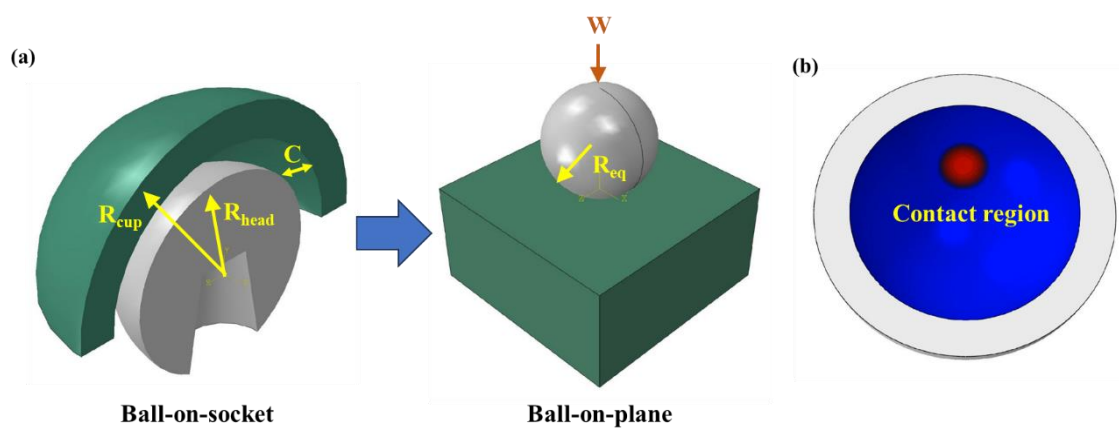

**Fig. 1.** a) Equivalent ball-on-plane model for the realistic ball-on-socket joint in hard-on-hard hip implants. The cross section is shown in the ball-on-socket to define the geometry b) Typical contact size in hard-on-hard implants at Paul peak load (3000 N) estimated using finite element method

The equivalent modulus $E_{eq}$ can be predicted as shown in the Eq.(3).

$$E_{eq} = 2\left[\frac{1-v_{head}^2}{E_{head}} + \frac{1-v_{cup}^2}{E_{cup}}\right]^{-1} \quad (3)$$

**2.3 Classification of Lubrication regime**

Minimum lubrication film thickness is calculated using the empirical Hamrock-Dowson formula [27] subjected to iso-viscous and elastic lubrication, as shown in Eq.(4) for the whole gait cycle. This equation holds for the equivalent ball-on-plane model under quasi-static conditions. The minimum film thickness ($h_{min}$) is predicted using Eq.(4) for a dynamic load variation in this study.

$$h_{min} = 2.8\, R_{eq} \left(\frac{\eta u}{E_{eq} R_{eq}}\right)^{0.65} \left(\frac{W}{E_{eq} R_{eq}^2}\right)^{-0.21} \quad (4)$$

Where $\eta$ is the viscosity of the body fluid (*Pa.s*), $W$ is the applied load (*N*), $u$ is the entrainment velocity (*m/s*).

Theoretical predictions of the lubrication regimes are due to a parameter specific film thickness $\lambda$, which will be calculated using the Eq. (5).

$$\lambda = \frac{h_{min}}{R_a'} \quad (5)$$

Where $R_a'$ is the composite roughness which is calculated by the root mean square of the individual surface roughness, i.e., $\sqrt{R_{a,head}^2 + R_{a,cup}^2}$. Based on the value of the $\lambda$, the regimes are classified [27] and shown in the Fig. 2.

$\lambda < 1$ : boundary lubrication

$1 \leq \lambda \leq 3$ : mixed lubrication

$\lambda > 3$ : full-film lubrication

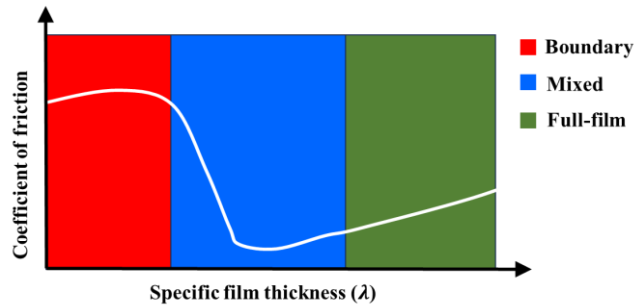

**Fig. 2.** Stribeck curve showing the classification of lubrication regimes

## 2.4 Loads, Rotations and Lubricant properties

As walking is one of the most common activities [29], the loading conditions used for the present study are taken from a normal walking gait cycle [27], as shown in Fig.3a. The gait pattern forces taken in 3 component directions along with their resultant are plotted in terms of body weight percentage so that different weight cases of patients can be investigated. From Fig.3a, it is observed that the force component along y-direction (Superior-Inferior Axis) is about 3.5 times the body weight, irrespective of the patient's individual weight. To examine the influence of body weight in the lubrication regime, four different body weights from normal to obese persons (60,80,100 and 120 kg) are considered in this study. The resultant force will be considered as the load applied on the ball described in Fig.1a.

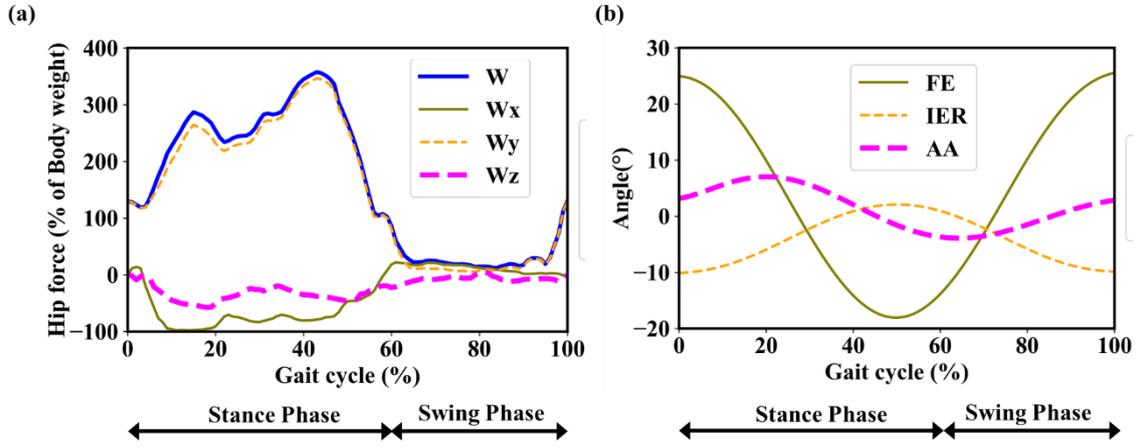

**Fig. 3.** a) Hip joint forces according to normal walking gait along 3 components with the resultant b) Angular displacements of the hip joint according to ISO 14242-1 which is similar to normal walking cycle

The rotations are taken from ISO 14242-1 which produces the angular displacements similar to a normal walking cycle. Fig.3b shows the angular displacements along three anatomical axes, namely Flexion-Extension (FE), Abduction-Adduction (AA) and Internal-External Rotation (IER). The mean cycle time is taken according to ISO standards as $1 \pm 0.1$ Hz. The slide track distances are determined by rotating the hip angles by the nodes in contact on the surface of the femoral head by the Eulerian rotation matrix M at discrete time instants according to FE→AA→IER as shown in Eq. (6).

$$M_i = \begin{bmatrix} sin\alpha_i sin\beta_i sin\gamma_i + cos\alpha_i cos\gamma_i & -sin\alpha_i cos\beta_i & sin\alpha_i sin\beta_i cos\gamma_i + cos\alpha_i sin\gamma_i \\ cos\alpha_i sin\beta_i sin\gamma_i + sin\alpha_i cos\gamma_i & cos\alpha_i cos\beta_i & -cos\alpha_i sin\beta_i cos\gamma_i + sin\alpha_i sin\gamma_i \\ -cos\beta_i sin\gamma_i & sin\beta_i & cos\beta_i cos\gamma_i \end{bmatrix} \quad (6)$$

Where $i$ represents the discrete time instant and $\alpha, \beta, \gamma$ are the angles of FE, AA and IER respectively.

Using the first-order forward and backward schemes at the extremes, the Finite Difference Method (FDM) is used to calculate the angular velocities at discrete time instants based on the angular displacements. Further, the second-order central difference scheme for the in-between values to calculate the angular velocities. As the acetabular cup is fixed to the pelvis bone ($V_{cup} = 0$) and the femoral head is the rotating component, the resultant relative velocity is calculated according to the head. The resultants of the obtained angular velocities are multiplied by the radius of the femoral head to acquire the resultant linear velocity $V$ for the head at the respective time instants as shown in Eq. (7).

$$V_i = \omega_i R_{head} \qquad (7)$$

Where $\omega_i$ is the resultant angular velocity of the head at discrete time instants.

The entrainment velocity $u$ at specific time instants is calculated for the tribo-pair as in Eq. (8).

$$u_i = \frac{V_i}{2} \qquad (8)$$

Studies have reported that after the hip replacement surgery, synovial fluid (SF) behaves as periprosthetic [27] which different from a healthy synovial fluid property. Ideally, the fluid has a non-Newtonian behaviour, but in this study, a simple Newtonian, incompressible and iso-viscous lubricant model is assumed. Typical viscosity value of 0.0025 Pa.s is taken for the periprosthetic SF in this paper.

**2.5 Hertzian point contact**

As discussed in section 2.2, Hertz theory of point contact which is based on the equivalent ball-on-plane model is considered to understand the impact of the parameters. Since the contact profile is a circle, as shown by Fig. 1b, suitable equations are taken into consideration. From the Hertzian theory of point contact, the contact radius $a$, and the maximum contact pressure $P_{max}$ relations are shown in Eq.(9) and Eq.(10) respectively.

$$a = \sqrt[3]{\frac{3WR_{eq}}{4E_{eq}}} \qquad (9)$$

$$P_{max} = \frac{3W}{2\pi a^2} \qquad (10)$$

## 3. Results and Discussion
### 3.1 Slide track profile and sliding velocity

According to the previous studies [30,31], a minimum time step discrete instants of 40 must be present to have a convergence of the slide track analysis. However, in the slide track analysis for a smooth profile generation, the entire angular displacement values over a single gait cycle as shown in Fig.3b are split into 100-time steps. Fig.4 shows the slide track profile covered by the nodes on the femoral head calculated using Eq. (6) for different head diameters irrespective of the clearance and the material combination. In general, the slide track profile looks similar to an "ellipse" which is reported by earlier studies [32–34] for the ISO 14242-1 the normal walking condition. It should be noted that the slide track profile increases in size as the diameter of the femoral head increases. This is because as the diameter of the femoral head increases, distance covered by the nodes in contact increases to cover the same motion caused by the angular displacement of the given gait cycle.

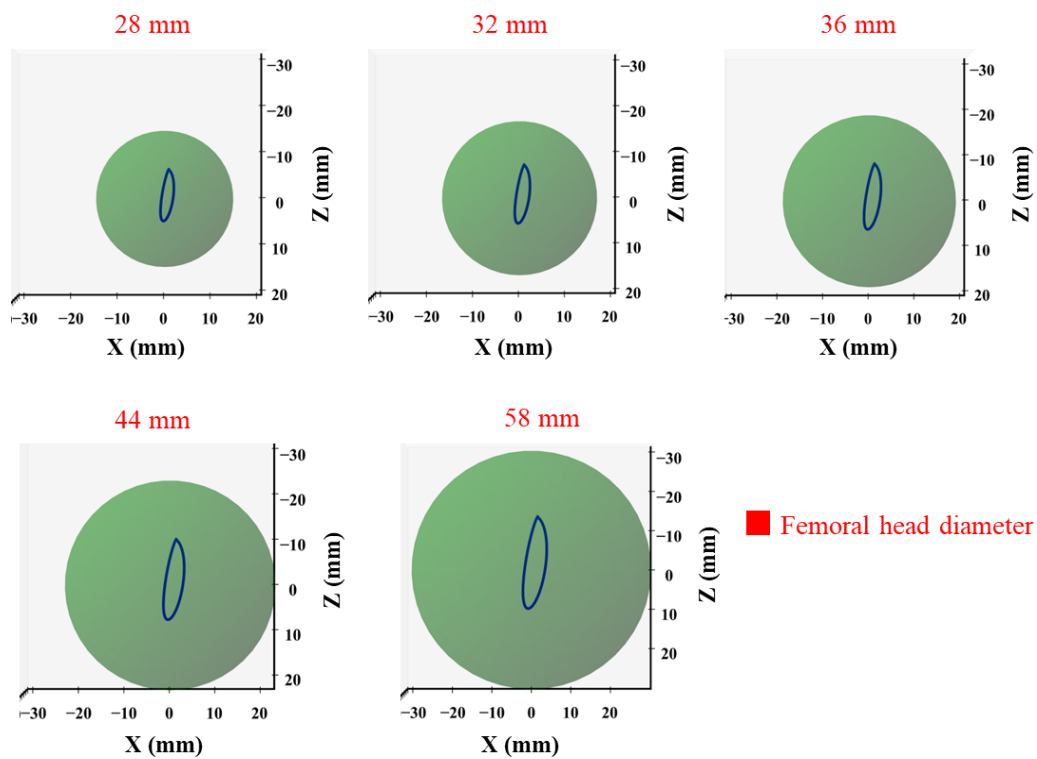

**Fig. 4.** Slide track profile covered by the nodes of the femoral head during a normal walking cycle with different head sizes

Further, the resultant linear velocity over the gait cycle is found using Eq.(7) for different head sizes irrespective of the clearance and the material combination as shown in Fig.5. It denotes the velocity with which the patient covers the particular gait cycle. It should be observed that for a given velocity profile, the magnitude of the resultant relative linear velocity grows with increasing femoral head size. A larger distance is travelled for the same gait cycle time due to the increase in femoral head diameter. The maximum resultant relative linear velocity occurs at the 25% and 75% of the gait cycle; whereas the minimum relative velocity occurs at 0% and 100% (start and stop condition) of the gait cycle. The velocity reaches to a very low value at 50% of the gait cycle which denotes there is a change in direction of motion in the ellipse track (i.e., major axis end).

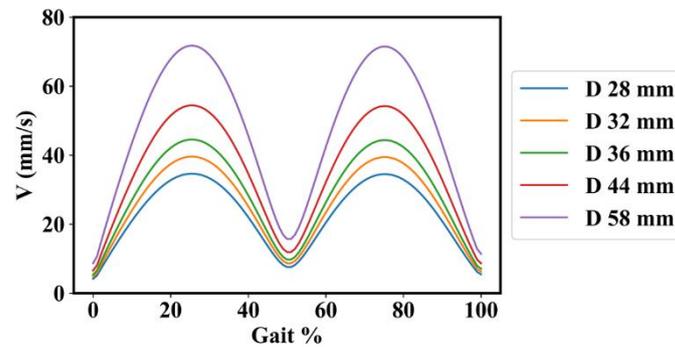

**Fig. 5.** Resultant relative velocity profile between the head and the cup for different femoral head sizes under normal walking gait ISO 14242-1

### 3.2 Minimum film thickness in MoM tribo-pair

Fig.6 shows the contour plot of minimum lubrication film thickness ($h_{min}$) within the whole gait cycle for the different femoral head sizes and body weights for corresponding radial clearances of 25, 50, 75 and 100 μm suitable for MoM. The effect of the individual parameters influencing $h_{min}$ will be described in the forthcoming sections.

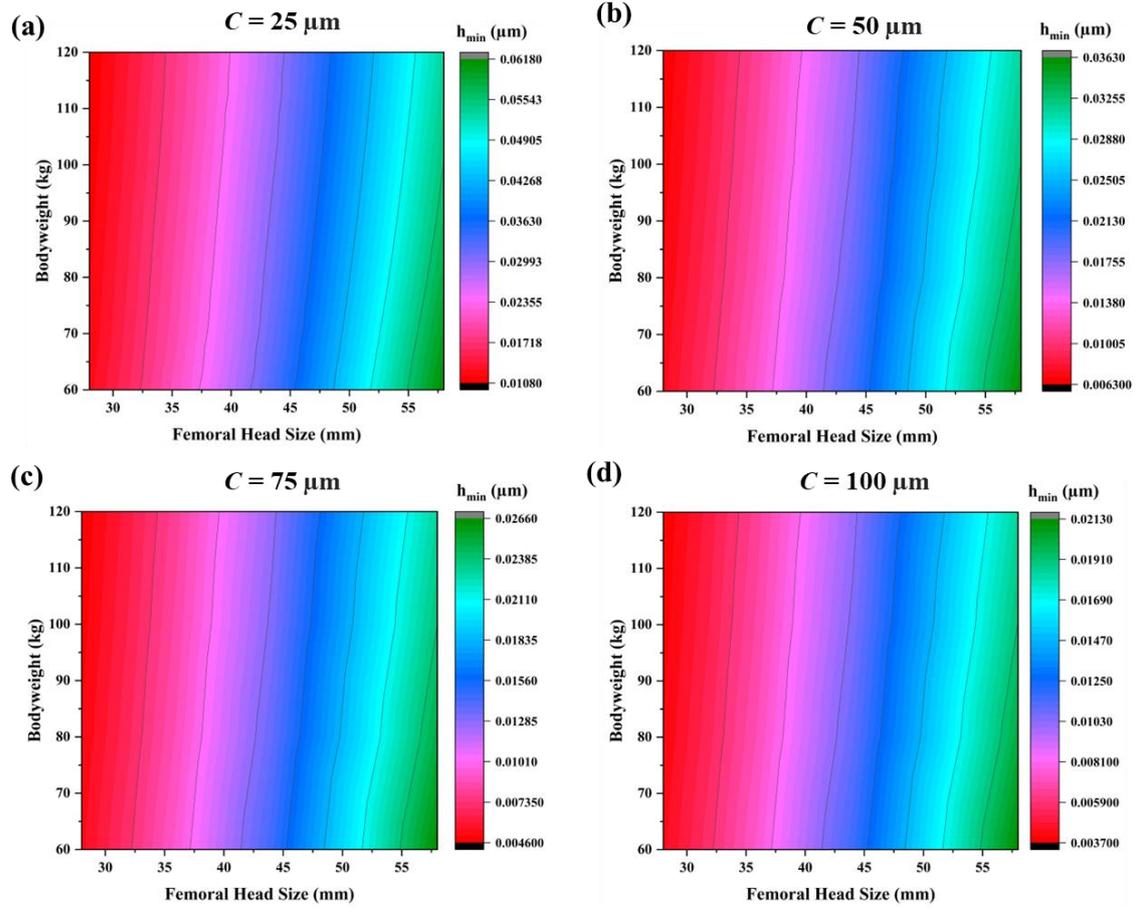

**Fig. 6.** Contour plot of $h_{min}$ in MoM within the whole gait cycle for the different femoral head sizes and body weight for radial clearances of a) 25 μm b) 50 μm c) 75 μm and d) 100 μm

### 3.2.1 Effect of Body weight

It is observed from Fig.6 that when the body weight increases starting from 60 to 120 kg, the $h_{min}$ decreases for a fixed femoral head size irrespective of the radial clearance C. The body weight parameter has an influence not only on contact pressure but also on the contact radius as described in Eq. (9) and Eq. (10), respectively. The increase in maximum contact pressure due to the applied load in the ball-on-plane model is high compared to the increase in the contact radius as inferred from the above equations in this section. As an effect, this increase in maximum contact pressure causes the thickness of the lubricant to reduce. Further, to study the impact of body weight, data points of the Fig.5 are plotted in Table 3 for a fixed clearance of 25 μm.

**Table 3**

$h_{min}$ values in μm for $C = 25$ μm

|  | $h_{min}$ (μm) | | | |
|---|---|---|---|---|
|  | **BW 60 kg** | **BW 80 kg** | **BW 100 kg** | **BW 120 kg** |
| **D 28 mm** | 0.112 | 0.105 | 0.101 | 0.096 |
| **D 32 mm** | 0.150 | 0.141 | 0.135 | 0.130 |
| **D 36 mm** | 0.194 | 0.183 | 0.174 | 0.167 |
| **D 44 mm** | 0.301 | 0.284 | 0.271 | 0.260 |
| **D 58 mm** | 0.551 | 0.519 | 0.495 | 0.477 |

From Table 3, it is perceived that the change in $h_{min}$ due to the body weight is less significant on the film thickness compared to the femoral head size. Similar trend is observed for other clearances with the change in body weight affecting $h_{min}$ as represented in Fig.6.

### 3.2.2 Effect of Femoral head size

Eq. (4) can be simplified to get the $h_{min}$ in terms of the femoral head size affecting terms by simplifying to Eq.11.

$$h_{min} = K R_{eq}^{0.77} u^{0.65} \tag{11}$$

Where $K$ refers to the remaining terms in Eq. (4).

The entrainment velocity in Eq. 11 can be expanded in terms of femoral head radius to get Eq.12,

$$h_{min} = K R_{eq}^{0.77} (R_{head})^{0.65} \left(\frac{\omega}{2}\right)^{0.65} \tag{12}$$

From Eq.12, it should be noted that for the same rotational conditions, the effect of the femoral head radius is influenced by the equivalent radius of curvature $R_{eq}$ and velocity terms. Also, it is observed from Fig.6 that the $h_{min}$ increases as the femoral head size increases for the same body weight irrespective of the radial clearance C. This is due to the increase in the equivalent radius of curvature $R_{eq}$ and entrainment velocity $u$ for the larger femoral head size for the ball-on-plane model. The phenomenon of change in $R_{eq}$ can be explained by the Hertz theory of point contact considered for the ball-on-plane configuration described in Eq. 9.

The increase in equivalent radius (more conformal) for the larger femoral head size causes the contact radius to increase, thereby reducing the pressure at the region of contact. This reduction in contact pressure, with increased contact area, causes the lubricant film to distribute well throughout the contact area. The smaller the equivalent radius of curvature causes a small contact radius, resulting in the maximum contact pressure to increase, signifying less conformal surfaces.

The impact of femoral head size is higher when compared to the change in body weight for a given clearance, which is also shown in Table 3. Further, it is also evident from Fig.6 to state that the maximum value of $h_{min}$ occurs at the least body weight of 60 kg and the largest femoral head size of 58 mm considered in the study, irrespective of clearance. The minimum value of $h_{min}$ occurs at the maximum body weight of 120 kg and the least femoral head diameter of 28 mm.

### 3.2.3 Effect of clearance

**Table 4**

Effect of radial clearance in affecting $h_{min}$ for MoM

| Clearance $C$ (μm) | Minimum film thickness $h_{min}$ range (μm) | Mean $h_{min}$ film thickness (μm) | Change in mean film thickness (%) |
|---|---|---|---|
| 25 | 0.0108 - 0.0618 | 0.0271 | - |
| 50 | 0.0063 - 0.0363 | 0.0159 | 41.32 |
| 75 | 0.0046 - 0.0266 | 0.0116 | 27.04 |
| 100 | 0.0037 - 0.0213 | 0.0093 | 19.82 |

An additional finding from Fig.6 is that when the value of radial clearance $C$ increases, the range of $h_{min}$ decreases. As shown in Table 4, the percentage change in the mean value of the $h_{min}$ is 41.35%, 26.84% and 19.42% for the shift in radial clearance from 25-50, 50-75 and 75-100 μm, respectively. The radial clearance has an influence in changing the equivalent radius of curvature in the $h_{min}$ calculation. Interestingly, for the smaller magnitudes of $C$ causes the $R_{eq}$ value to drastically increase as described in Eq.2. Since, the impact of changing small radial clearances is more significant in affecting $h_{min}$ which is evident from Table 4, only 25 μm and 50 μm clearances are considered for further analysis in this study.

### 3.2.4 Effect of gait pattern force and articulating velocity

The gait pattern force refers to the specific forces generated at the joint due to the type of gait movement such as walking, staircase climbing etc., considered for motion. It is expressed as a percentage in terms of body weight as in Fig. 3a. To see the evolution of gait pattern load and articulating velocity, $h_{min}$ is calculated for the whole cycle calculated for different femoral head sizes, clearances and body weights as shown in Fig. 7. Since the percentage change in $h_{min}$ is high (≈ 40%) in between the two clearances 25 μm and 50 μm as shown in Table 4, only these clearances are considered as critical for analysing $h_{min}$ throughout the cycle. From Fig. 7 (in all plots), it is deduced that the effect of body weight is less significant which is discussed in section 3.2.1. Nevertheless, it should be highlighted that the value of the $h_{min}$ remains high during the swing phase of the gait cycle compared with the stance phase for all head sizes and clearances including body weights as well. This is due to the constant along with less swing phase gait load compared to the varying high magnitude stance phase gait load in the cycle. From the plots, it is observed that the type of gait pattern has a significantly influence on the $h_{min}$ compared to the body weight.

Further, the $h_{min}$ value is minimum at the points where the resultant velocity remains minimum, as discussed in section 3.1. This is because of the start-stop conditions occurring in a single cycle, where the depletion of the lubricant occurs in the contact region. This phenomenon is observed for both hard-on-hard and hard-on-soft combinations using hip simulator studies as reported [35]. However, the $h_{min}$ value is noticeably high at other places in comparison to the start-stop time instants. More lubricant is pulled into the contact surfaces due to the increase in relative velocity between the surfaces [36]. This indicates that the minimal film thickness is also significantly influenced by the articulating velocity.

As discussed in section 3.2.2, when the head size increases the value of $h_{min}$ increases for the same body weight and radial clearance C. Furthermore, the change in $h_{min}$ parameter is very high (≈ 40 % reduction) for the change in clearance $C$, for the same head size. Moroever, the effect of the individual parameters affecting $h_{min}$ will be studied for CoC in the subsequent sections.

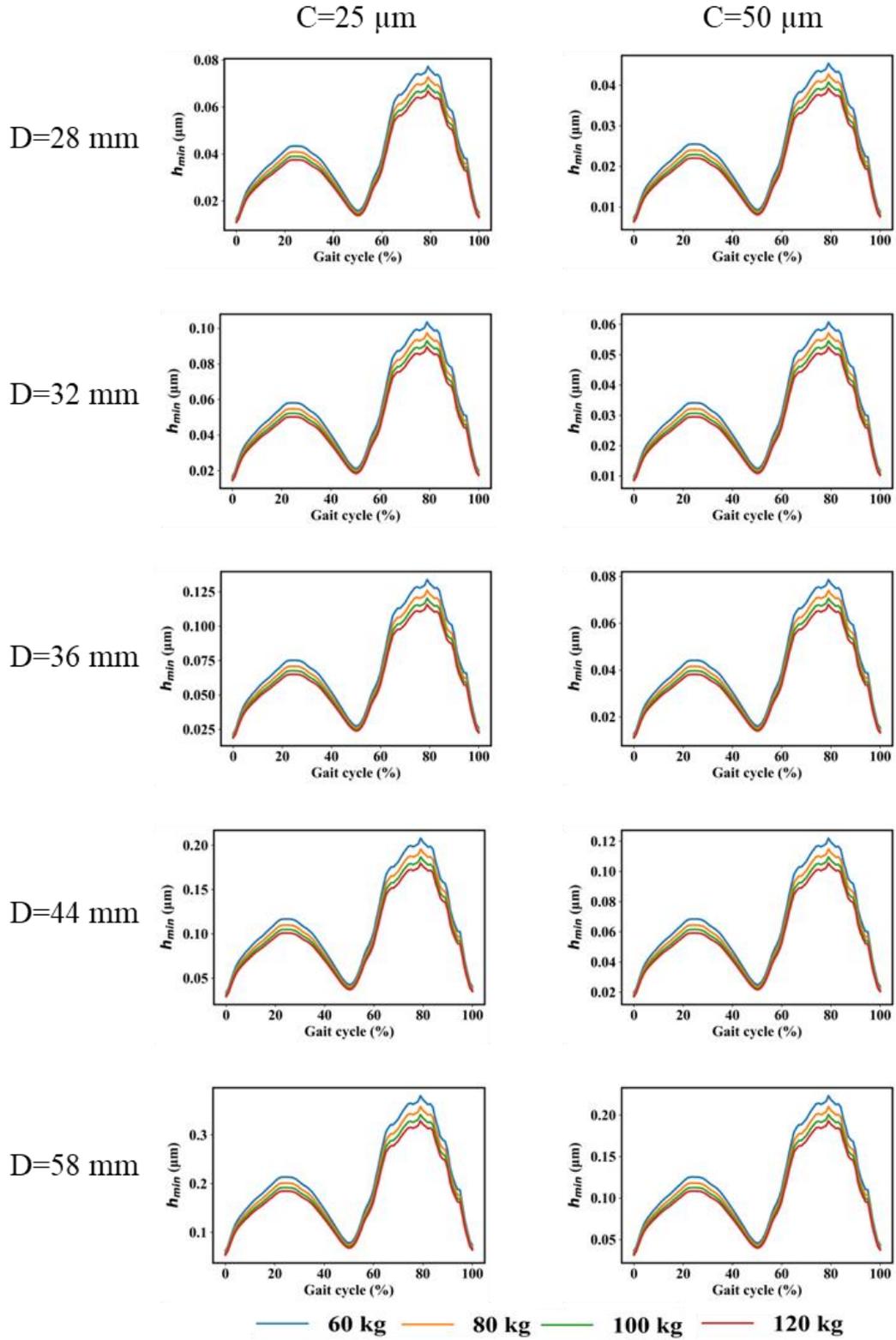

**Fig. 7.** $h_{min}$ values calculated for the whole gait cycle for different femoral head sizes, clearance range with respect to different body weights in MoM

## 3.3 Minimum film thickness – for CoC comparison

Fig.8 shows the contour plot of minimum lubrication film thickness ($h_{min}$) within the whole gait cycle for the different femoral head sizes and body weight for corresponding radial clearances of 25, 50, 75 and 100 μm for CoC. All four clearances, from 25 to 100 μm, are taken into account when assessing $h_{min}$ for conformance in CoC in order to examine how the selected parameters affect $h_{min}$ in relation to MoM.

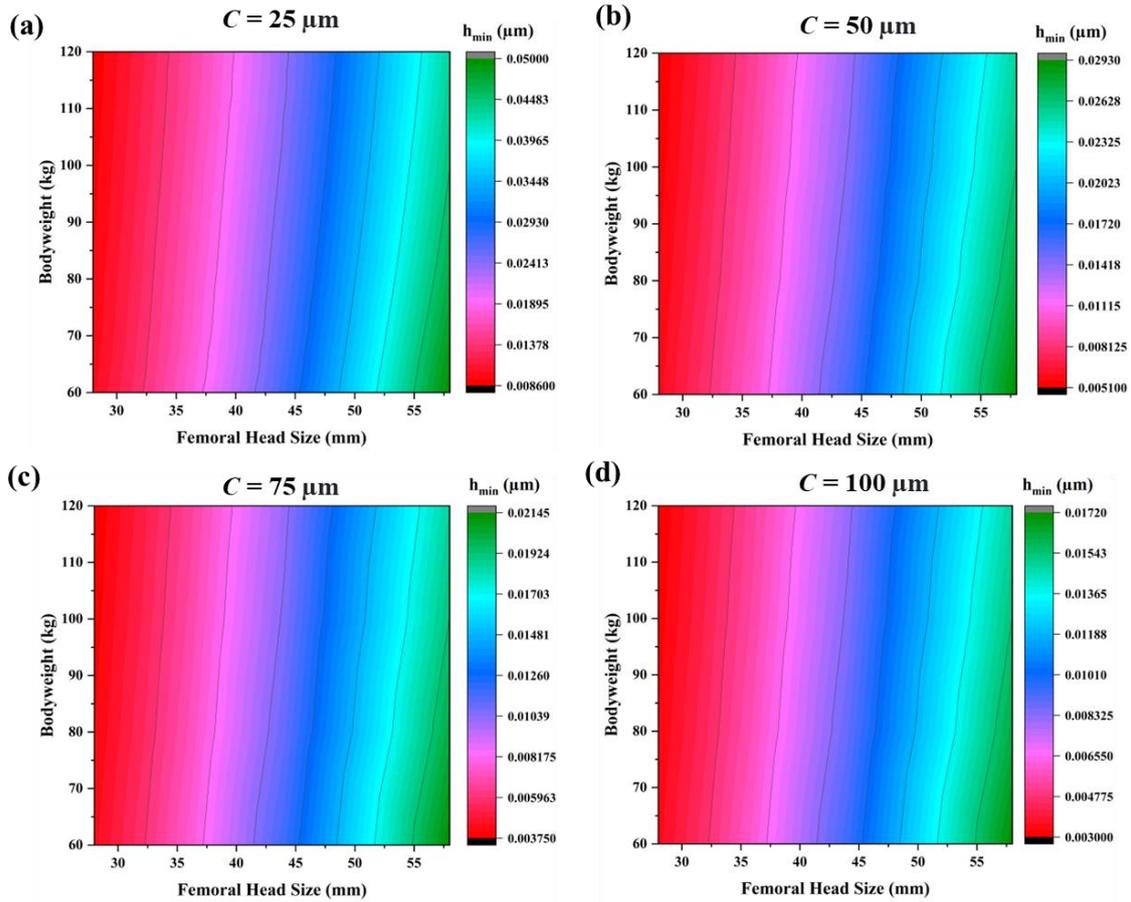

**Fig. 8.** Contour plot of $h_{min}$ in CoC within the whole gait cycle for the different femoral head sizes and body weight for radial clearances of a) 25 μm b) 50 μm c) 75 μm and d) 100 μm

### 3.3.1 Effect of the selected parameters in CoC

Comparing Figs.6 and 8, the effect of body weight, femoral head size and clearance in affecting minimum film thickness ($h_{min}$) are similar for both MoM and CoC tribo-pairs. The percentage change in $h_{min}$ due to clearance $C$ is, also, similar to MoM as represented in Table 4 and Table 5.

Table 5

Effect of radial clearance in affecting $h_{min}$ for CoC

| Clearance $C$ (μm) | Minimum film thickness $h_{min}$ range (μm) | Mean $h_{min}$ film thickness (μm) | Change in mean film thickness (%) |
|---|---|---|---|
| 25 | 0.0086-0.0500 | 0.0219 | - |
| 50 | 0.0051-0.0293 | 0.0128 | 41.55 |
| 75 | 0.0038-0.0215 | 0.0094 | 26.56 |
| 100 | 0.0030-0.0172 | 0.0075 | 20.21 |

The profile of $h_{min}$ due to gait pattern load and angular velocity is, also, similar to MoM tribo-pair but the value of $h_{min}$ is substantially less as compared to MoM as shown in Fig.9.

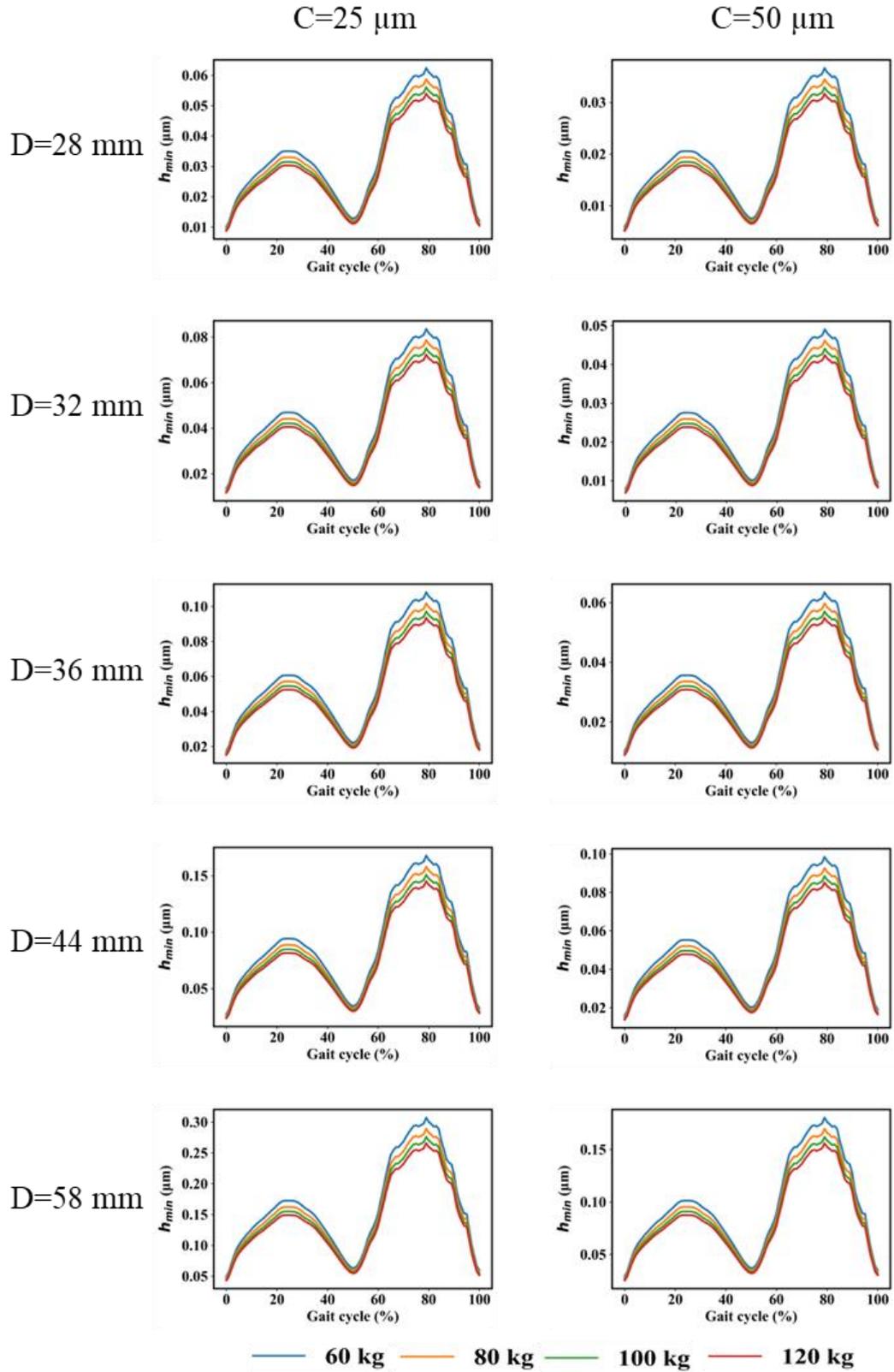

**Fig. 9.** $h_{min}$ values calculated for the whole gait cycle for different femoral head sizes, clearance range with respect to different body weights in CoC

### 3.4 Effect of Young's modulus and Poisson's ratio between MoM and CoC

Fig.10 shows that the magnitude of $h_{min}$ is less for CoC compared to MoM for all the conditions with the effect of body weight, femoral head size and clearance $C$.

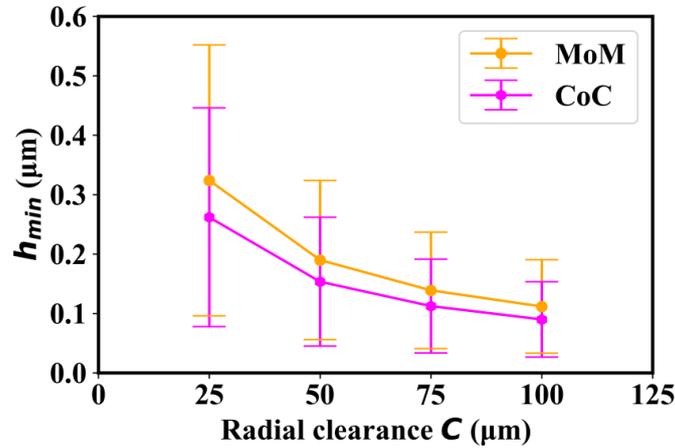

**Fig. 10.** Effect of radial clearance C (mean value with deviation) in the $h_{min}$ for MoM and CoC

As ceramics have a higher Young's modulus than metals, scientifically the equivalent modulus $E_{eq}$ of the material increases according to Eq. (9) which leads to a decrease in the contact radius (more non-conformal contact). In a ball-on-plane model, this reduction in contact radius causes an increase in pressure, as described by Eq. (10), which lowers the $h_{min}$ in CoC hip implants. Further, it is also observed that the deviation of $h_{min}$ decreases along with the increase in radial clearance $C$ for both MoM and CoC.

### 3.5 Quantitative effect of individual parameters affecting $h_{min}$

Until now, the influence of the individual parameters affecting the $h_{min}$ is studied qualitatively in the preceding sections. It is important to remember that a couple of parameters are adjusted incrementally based on the author's interests, while other parameters (head, material, and loading standards) are based on implants manufacturers catalogue specifications. Therefore, the Pearson correlation coefficient ($r$) is computed with a range of parameters in order to examine the quantitative impact and normalisation of each parameter on the $h_{min}$ value, as indicated in Table 6. The gait pattern forces are considered between the range 1.76 to 34.33 *N/kg* based on the normal waking gait as represented in Fig.3a. The angular velocities, which are constant irrespective of the head size are considered according to ISO 14242-1 and are taken in the range between 0.3-2.5 *rad/s*. As only two materials are considered for hard-on-hard combination, another biocompatible material of a different Young's modulus and Poisson's ratio [37], i.e., alumina ($E$ = 413 GPa and $v$ = 0.235) is considered to capture the trend of material properties.

**Table 6**

Parameters considered for determining correlation coefficients

| Materials | CoCrMo, ZTA, Alumina |
|---|---|
| Femoral head diameter, *D* (*mm*) | 28, 32, 36, 44, 58 |
| Radial clearance, *C* (*μm*) | 25, 50, 75, 100 |
| Body weight, *BW* (*kg*) | 60, 80, 100, 120 |
| Viscosity of the body fluid, (*Pa.s*) | 0.0025 |
| Angular velocity, (*rad/s*) | 0.5, 1.0, 1.5, 2.0, 2.5 |
| Gait pattern force, (*N/kg*) | 4.905, 9.810, 14.715, 19.620, 24.525, 29.430, 34.335 |

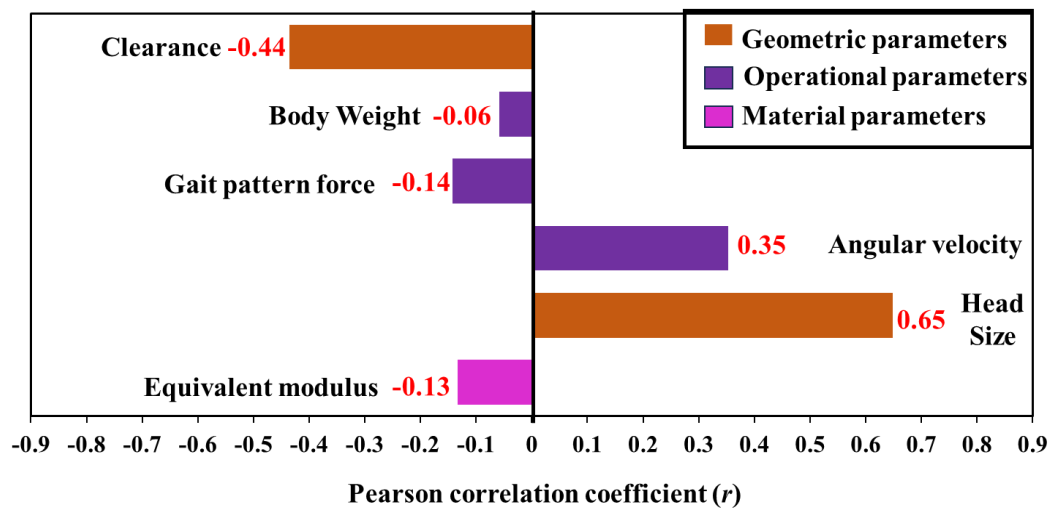

**Fig. 11.** Correlation coefficients generated for individual parameters affecting $h_{min}$ considered in the study

Fig. 11 shows the parameters and corresponding correlation coefficients that are considered for the purpose of this study. It is observed that angular velocity and head size are positively correlated with the $h_{min}$ whereas clearance, body weight, gait pattern force and equivalent modulus are negatively correlated. Even though the effect of clearance appears to be qualitatively higher than the change in head size as seen in Figs. 7 and 9, contrastingly the head size (r = 0.65) has a greater correlation than the clearance (r = -0.44) among the geometric parameters.

The angular velocity between the cup and head has the highest correlation among the operational parameters on $h_{min}$. Further, among the operational parameters, it is quite clearly observed that the body weight ($r = -0.06$) has the least effect. In the remaining material parameters, the type of material tribo-pair significantly affects $h_{min}$. The stiffer the material is, the thickness of the lubricant becomes less.

From the correlation coefficients, the parameters are ranked in the descending order as head size, clearance, angular velocity, gait pattern force, equivalent modulus and the body weight. Overall, the geometrical parameters have the maximum influence on $h_{min}$ followed by operational and material parameters. As this study involves only the gait conditions suggested by ISO 14242-1(normal walking), the operational parameters due to normal walking gait has less significance on $h_{min}$ compared to the geometrical parameters. At the same time, due to other realistic loading conditions i.e, staircase climbing, descending, sit ups etc., the operational parameters may have more significant effects on $h_{min}$. This study is focused on the parameters that would help to orthopaedist to select an appropriate implant prior to hip replacement.

With the view of an orthopaedic surgeon regarding the selected parameters as controllable or uncontrollable, as discussed in the introduction, it is evident that the controllable parameters (head size and clearance) have a more dominant effect on $h_{min}$. Given that the angular velocity and gait pattern force also have significant effects on $h_{min}$, it implies the characteristic of the individual patient. The surgeon can only suggest the type of gait movement to be followed, but if the patient performs any extreme gait activities, consequently it may lead to a reduction in $h_{min}$. In general, the angular velocity between the tribo-pair decreases with the increase in the patient's age. So, the elderly patients have a slower gait movement, causing a reduction in $h_{min}$ value. As a result, the focus should be given on the controllable parameters than uncontrollable. Among the controllable parameters, choosing a larger head size provides stability, proper load distribution and improved range of motion for the patient [38,39]. A large group of people are migrating towards the larger head sizes for improved activities as reported which is also supported by this study [25]. Therefore, it is recommended that the surgeon should select a larger head size with less clearance which is suitable for the longevity of the hip implant.

**3.6 Specific film thickness and lubrication regime classification**

To examine the influence of surface roughness on the materials and to determine the lubrication regime, specific film thickness is calculated using Eq.(5) for both MoM and CoC tribo-pairs. A mean roughness value $R_a$ of 0.03 μm and 0.003 μm are taken for MoM and CoC respectively as reported [28]. Fig.12 shows the lubrication regimes within a single gait cycle derived from Fig.7 for different femoral head sizes, body weights and clearances of 25 µm and 50 µm in MoM. It should be noted that a single lubrication regime doesn't exist for a particular design hip implant due to the dynamically varying load and velocity of the gait cycle.

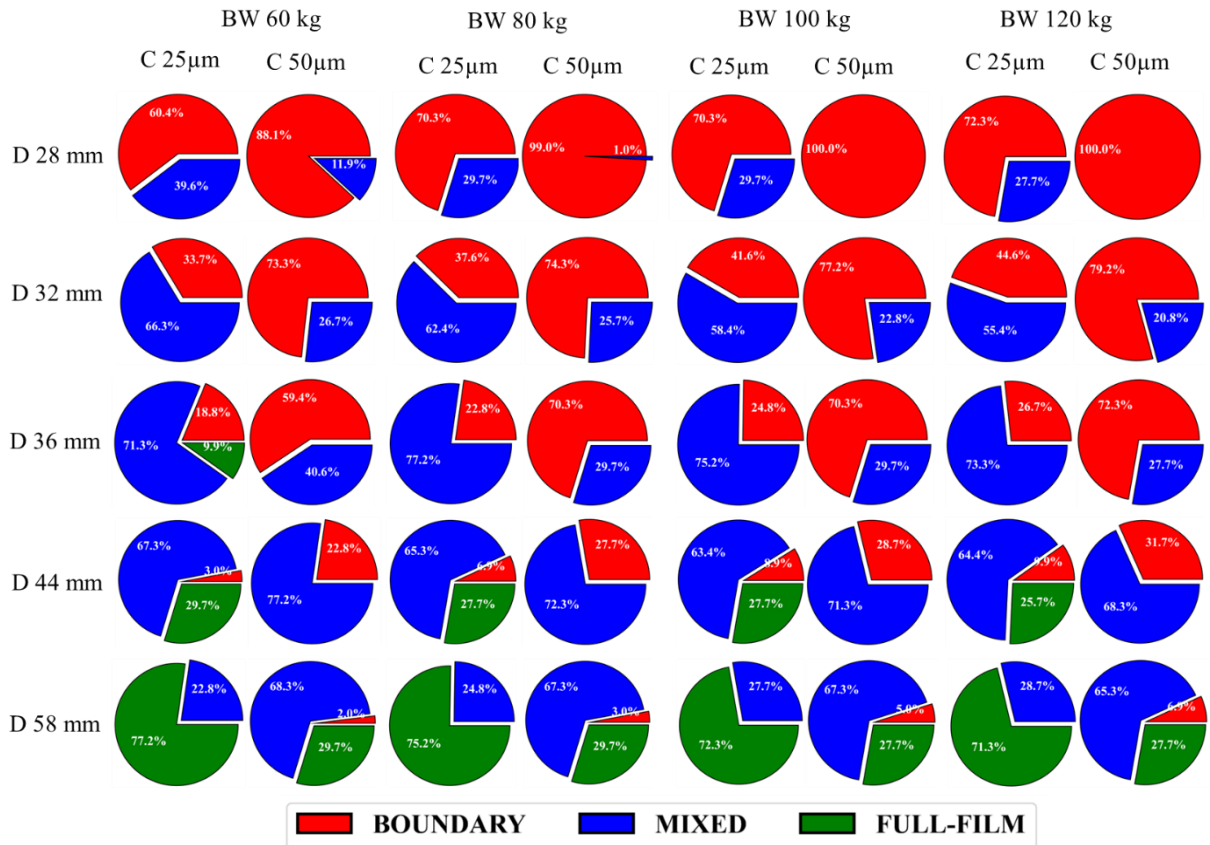

**Fig. 12.** Lubrication regimes classified within a single gait cycle of MoM tribo-pair according to $\lambda$ for different femoral head sizes, body weights and clearances of 25 µm and 50 µm

Most of the lubrication regime lies in mixed-boundary lubrication for the THR head sizes and a combination of boundary, mixed and full-film lubrication for the HRR head sizes for MoM tribo-pair. This observation of lubrication regime for the THR and HRR head sizes along with the $\lambda$ values is also reported for MoM [28]. At the same time, Fig.12 also infers that a large femoral head size with less clearance is preferred for higher tendency of the implant to perform in the full-film lubrication regime.

A similar analysis to MoM is performed on CoC tribo-pair to determine the lubrication regimes as represented in Fig.13. As discussed in section 3.4, given the same head size, body weight and clearance values, the value of minimum film thickness $h_{min}$ is less for CoC compared to MoM. In contrast, the value of $\lambda$ is very high (almost 10 times) in CoC compared to MoM causing the regime to shift towards full-film compared to the boundary-mixed lubrication regime of MoM. This phenomenon is probably due to the fine surface finish caused by the manufacturing process of ceramics but need to be confirmed.

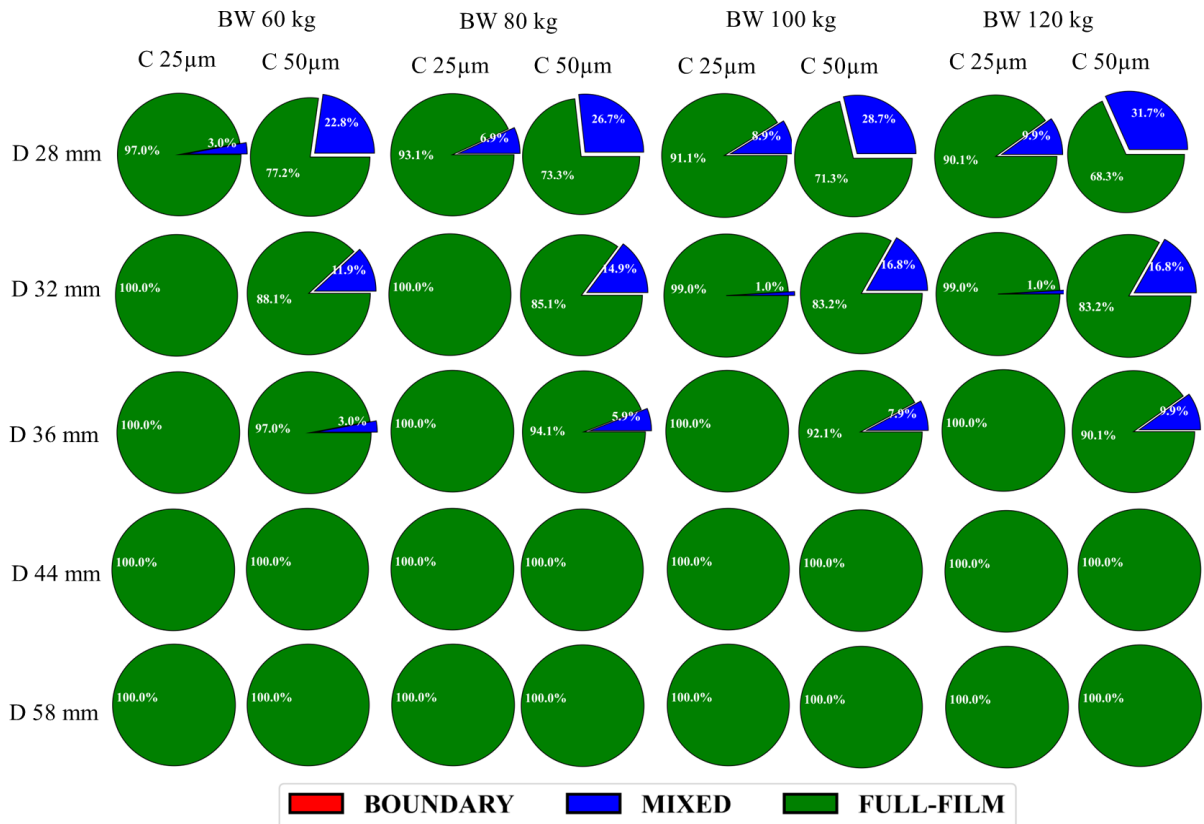

**Fig. 13.** Lubrication regimes classified within a single gait cycle of CoC tribo-pair according to $\lambda$ for different femoral head sizes, body weights and clearances of 25 µm and 50 µm

### 3.6.1 Effect of Surface Roughness $R_a$

Based on the results in the lubrication regime from Figs. 12 and 13, it is evident that CoC bearings perform better with respect to MoM. The manufacturing processes used in powder metallurgy enable ceramics with extremely tiny grain sizes; when the under with excellent polishing, produce a smooth surface finish. [40]. This smooth surface finish combined with wettability offers ceramics a better performance in lubrication and wear resistance than metals in joint applications [2,28,41]. From Fig.13, it is perceived that most of the lubrication regime in CoC tribo-pair lies predominantly in the full-film regime which is quite opposite to MoM. Full-film lubrication is 100% for all conditions in HRR head sizes for CoC tribo-pair. A different investigation from literature [28] also discovered the same full-film lubricating regime results for CoC hip implants.

It is also stated that the surface roughness of femoral head increases after implantation due to several mechanical and environmental factors in both hard-on-hard and hard-on-soft combinations [42,43]. In particular, the local surface roughness increases further in the presence of dislocation and impingement. Consequently, this results in a shift in the lubrication regime towards the boundary and, followed by excessive wear [44,45].

Moreover, the experimental hip simulator tests are performed without dislocation or impingement for the MoM tribo-pair [46]. It is observed that if the cup or head initially has a rougher surface than the counter surface, the rough surface becomes polished as the test duration increases, also increasing the surface roughness for the counter surface and producing similar values in the contact region. Thus, a smooth surface finish is preferred between the articulating cup and head surfaces. It is also found that the retrieved alumina femoral heads show the low surface roughness compared to Co-Cr metallic femoral heads [47]. Even though, the mechanical and surgical factors contribute to change in surface roughness, ceramics shows an improved full-film lubrication due to their ultra fine surface roughness which is also confirmed in this study through analytical formulations.

Though this study doesn't consider the effect of wettability between the contact surfaces, it is interesting to note that ceramics are hydrophilic compared to metals [40]. This leads to improved lubricant retention between the contact surfaces with uniform synovial fluid distribution. Also, low surface roughness provides better wettability leading to less friction and wear [48].

Therefore, it can be inferred from this study that an implant with a smooth surface is favoured because implant wear may be reduced in-vivo due to the prolonged evolution from the full-film to the boundary lubrication regime. As the CoC tribo-pair, also, has high wear resistance and bio-inert wear debris in addition to a fine surface finish, it is preferred over the MoM tribo-pair.

## 4. Conclusion

In the present study, the influence of geometrical parameters (femoral head size, clearance and surface roughness), material parameters (Young's modulus and Poisson's ratio) and operational parameters (body weight, gait pattern force and angular velocity) are studied, affecting the lubricating performance of hard-on-hard hip implants. The following conclusions are obtained from this study:

- Geometrical parameters are more significant in affecting the minimum film thickness followed by the operational and material parameters. Femoral head size is found to be the most dominant parameter and Body weight is found to be the least dominant parameter.

- Among the controllable parameters, choosing larger head size provides stability, proper load distribution and improved range of motion for the patient.
- For the given conditions, despite the $h_{min}$ of CoC being less than MoM tribo-pair, CoC performs predominantly in the full-film lubrication regime due to its ultra-fine surface finish.
- The present study supports that the orthopaedic surgeon, irrespective of any tribo-pair, should select the hip implant having a larger femoral head diameter, less clearance with ultra-fine surface finish to extend the hip implant lifespan.

**CRediT authorship contribution statement**

**K.Nitish Prasad:** Conceptualization, Methodology, Software, Visualization, Writing – original draft.

**P.Ramkumar:** Conceptualization, Investigation, Visualization, Supervision, Writing – review & editing.

**Declaration of competing interest**

The authors declare that they have no known competing financial interests or personal relationships that could have appeared to influence the work reported in this paper.

**Funding**

This research did not receive any specific grant from funding agencies in the public, commercial, or not-for-profit sectors.

**ORCID ID**

K. Nitish Prasad: https://orcid.org/0000-0001-9906-4124

P. Ramkumar: http://orcid.org/0000-0002-2816-9145

https://doi.org/10.1243/0954411001535228.

[47] P. Taddei, S. Tozzi, S. Carmignato, S. Affatato, May the surface roughness of the retrieved femoral head influence the wear behavior of the polyethylene liner?, J. Biomed. Mater. Res. - Part B Appl. Biomater. 104 (2016) 1374–1385. https://doi.org/10.1002/jbm.b.33483.

[48] S. Ghosh, S. Abanteriba, Status of surface modification techniques for artificial hip implants, Sci. Technol. Adv. Mater. 17 (2016) 715–735. https://doi.org/10.1080/14686996.2016.1240575.
31